\documentclass{ws-p8-50x6-01}

\newcommand{\beq}{\begin{equation}}
\newcommand{\eeq}{\end{equation}}
\newcommand{\bef}{\begin{figure}}
\newcommand{\enf}{\end{figure}}
\newcommand{\bec}{\begin{center}}
\newcommand{\enc}{\end{center}}
\newcommand{\bei}{\end{itemize}}
\newcommand{\eni}{\end{itemize}}

\newcommand{\mic}{$\mu$m }
\newcommand{\ingr}{\includegraphics}

\begin{document}

\title{Active controls in interferometric detectors of gravitational waves:
inertial damping of the VIRGO superattenuator\footnote{Lecture
given at the {\it International Summer School on Experimental
Physics of Gravitational Waves} - Urbino (Italy), September 6-18,
1999}}

\author{Giovanni Losurdo}

\address{Istituto Nazionale di Fisica Nucleare - Sezione di Pisa\\
E-mail: losurdo@galileo.pi.infn.it}


\maketitle

\abstracts{The operation of an interferometer for gravitational
waves detection requires sophisticated feedback controls in many
parts of the apparatus. The aim of this lecture is to introduce
the types of problems to be faced in this line of research. The
attention is focused on the {\it inertial damping} of the test
mass suspension of the VIRGO interferometer (the superattenuator):
it is a multidimensional local control aimed to reduce the
residual motion of the suspended mirror associated to the normal
modes of the suspension. Its performance is very important for the
locking of the interferometer. }

\section{Introduction}

Operating an interferometer for gravitational waves detection
requires the implementation of many active controls on the
different parts of the apparatus. For instance, feedback controls
are needed to reduce the laser frequency and power fluctuations,
to keep the optical cavities in resonance and to maintain the
interferometer output on a dark fringe. The basic idea is that the
apparatus works at its best strictly around a well defined working
point \cite{saulson} (laser fluctuations and shot noise at a
minimum, optical cavities in resonance and interferometer output
on a dark fringe). Internal and external disturbances overload the
dynamic range of the interferometer. Therefore, the system has to
be forced to remain in the correct working position. This is
achieved via feedback controls.

In this lecture we examine in detail one of the controls needed to
reach the final goal of operating the interferometer: the {\it
inertial damping} of the VIRGO superattenuator (SA) \cite{sa}. The
SA can be described as a chain of mechanical ``filters", each one
acting as a ``spring" in 6 degrees of freedom. The normal mode
frequencies of the SA range between 0.04 and about 2 Hz. At
frequencies $f>>2$ Hz the SA acts as a steep filter of the seismic
vibrations of the ground (an attenuation factor of $10^{15}$@10 Hz
is expected). Therefore, in the interferometer detection band (10
Hz-few kHz), the suspended mirror is ``disconnected" from the
ground. Beside being a low pass filter for ground vibrations the
SA is a tool for actively controlling the mirror position. Forces
that move and steer the mirror can be exerted in 3 points of the
SA chain: the inverted pendulum (IP) \cite{ip}, the {\it
marionetta} (a special mechanical tool designed to steer the
mirror) and the mirror itself (from a suspended {\it reference
mass}).

Keeping the interferometer in the operating position requires the
mirror to have a maximum RMS relative motion of less the
$10^{-12}$ m. In the frequency range where the SA is fully
effective ($f>>4$ Hz) the residual motion of the mirror is
negligible. On the other hand, the mirror free motion in the
region of the SA normal modes is $\sim 100$ \mic. Feedback forces
acting on the SA must reduce the mirror motion from $100$ \mic to
$10^{-12}$ m. The dynamic range of a feedback system able to
perform this control has to be huge: the control is performed in
three steps (hierarchical control \cite{amaldi97}). The first step
is a damping of the SA normal modes in order to reduce the mirror
residual motion to less than 10 \mic. This is necessary in order
to control the mirror position acting on the lower stages without
reinjecting noise into the detection band. We describe in the
following an implementation of a high gain and wideband damping of
the SA resonances.

\section{Experimental setup}

The setup (fig. \ref{setup}) of the experiment is composed of a
full scale superattenuator, provided with 3 accelerometers (placed
on the top of the IP), 3 LVDT position sensors (measuring the
relative motion of the IP with respect to an external frame), 3
coil-magnet actuators. The accelerometers work in the range DC-400
Hz and have acceleration spectral sensitivity $\sim 10^{-9}\; {\rm
m\,s}^{-2}\,{\rm Hz}^{-1/2}$ below 3 Hz \cite{acc}. The sensors
and actuators are all placed in {\it pin-wheel} configuration. The
sensor and actuator signals are computer controlled by a ADC (16
bit)-DSP-DAC (20 bit) system. The DSP handles the signals of all
the sensors and actuators. It can combine them by means of
matrices, create complex feedback filters (like the one of fig.
\ref{filter}) with high precision poles/zeroes placement and
perform all the calculations at a high sampling rate (10 kHz). The
suspended mirror is also provided with LVDT position sensors to
measure its displacement with respect to ground. \bef[t]\bec
\begin{minipage}{0.4\textwidth}
    \includegraphics[bb=2.5cm 1.5cm 17cm
28cm,clip=true,width=\textwidth]{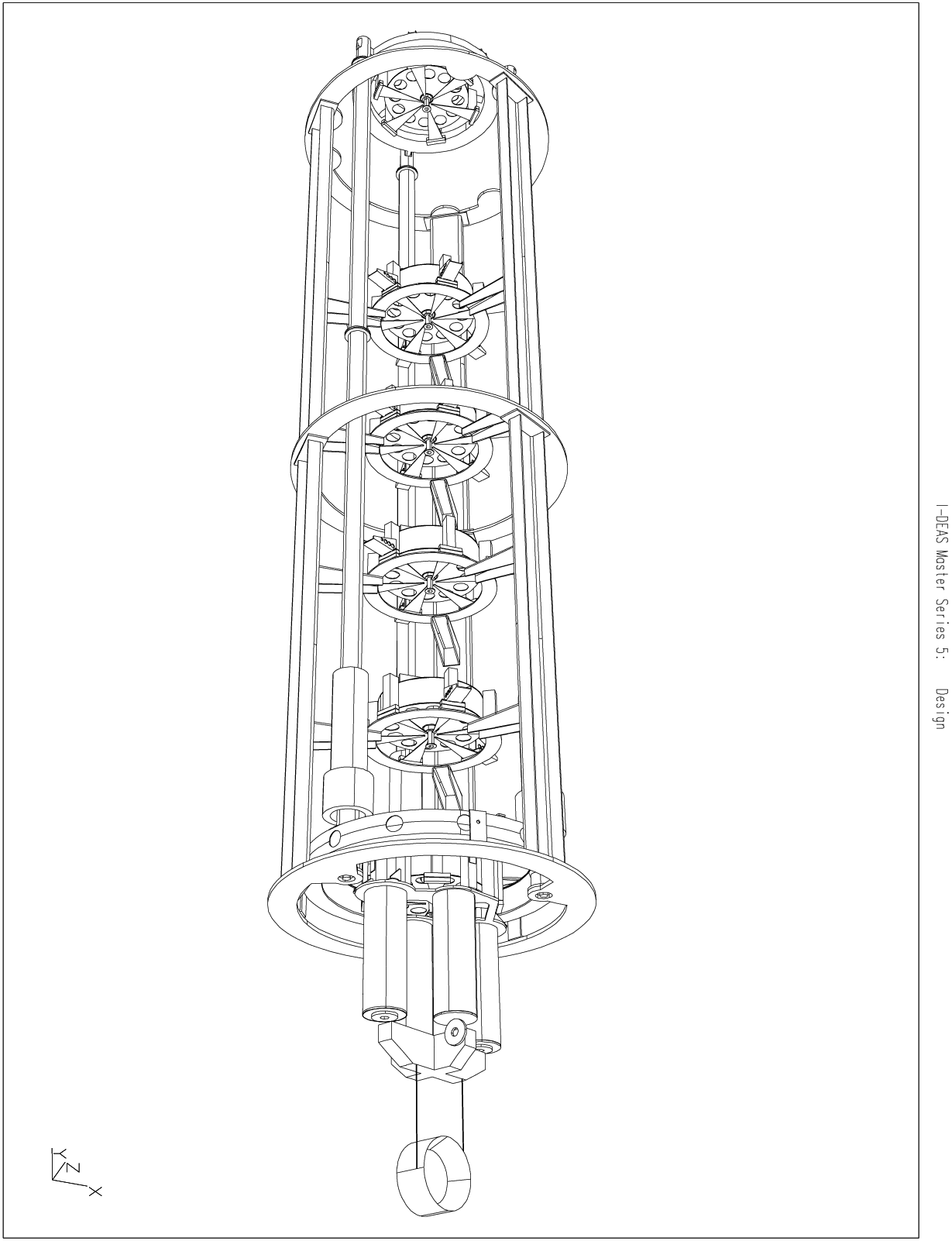}
\end{minipage}\hfill
\begin{minipage}{0.6\textwidth}
 \includegraphics[width=.9\textwidth]{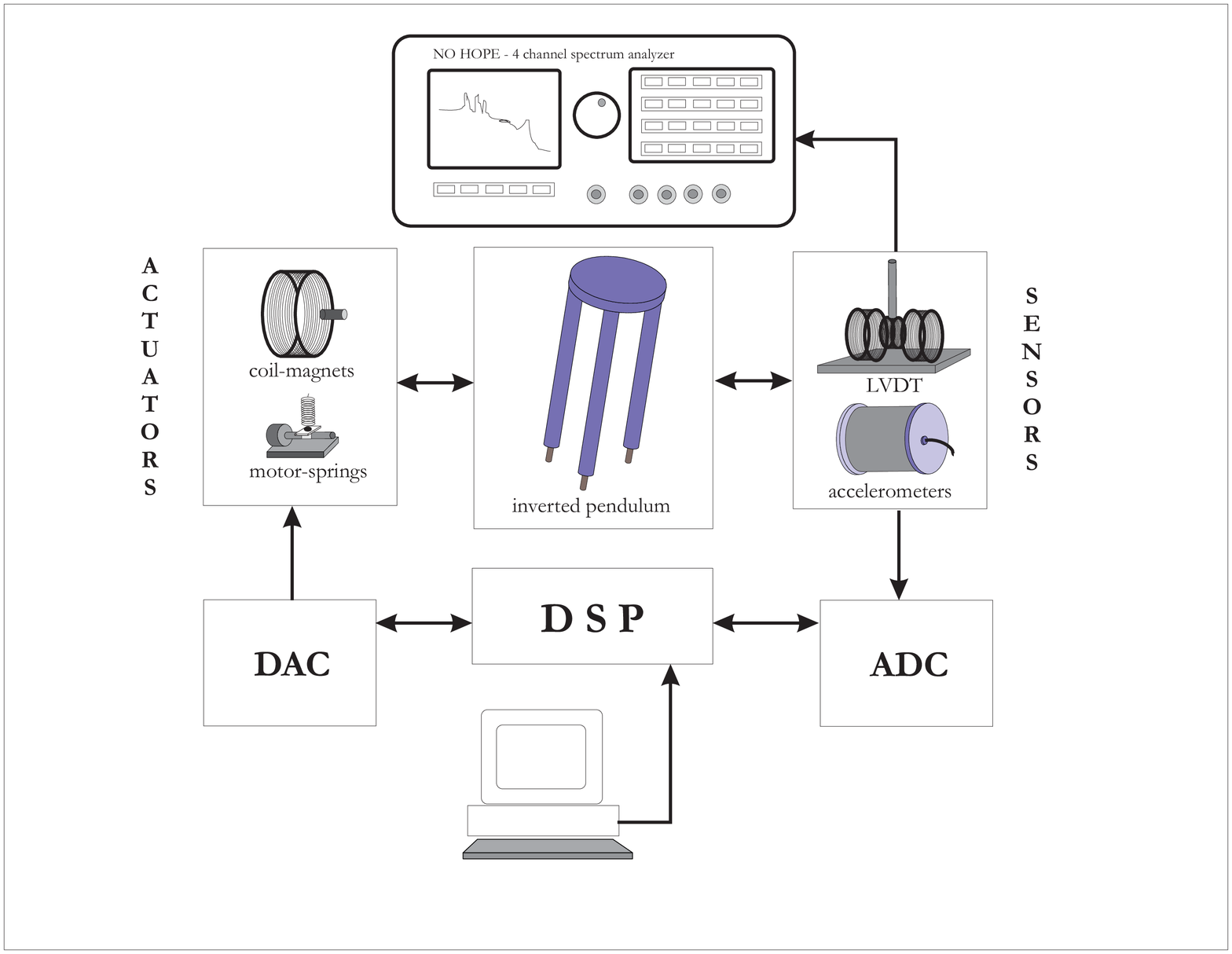}
 \includegraphics[width=0.95\textwidth]{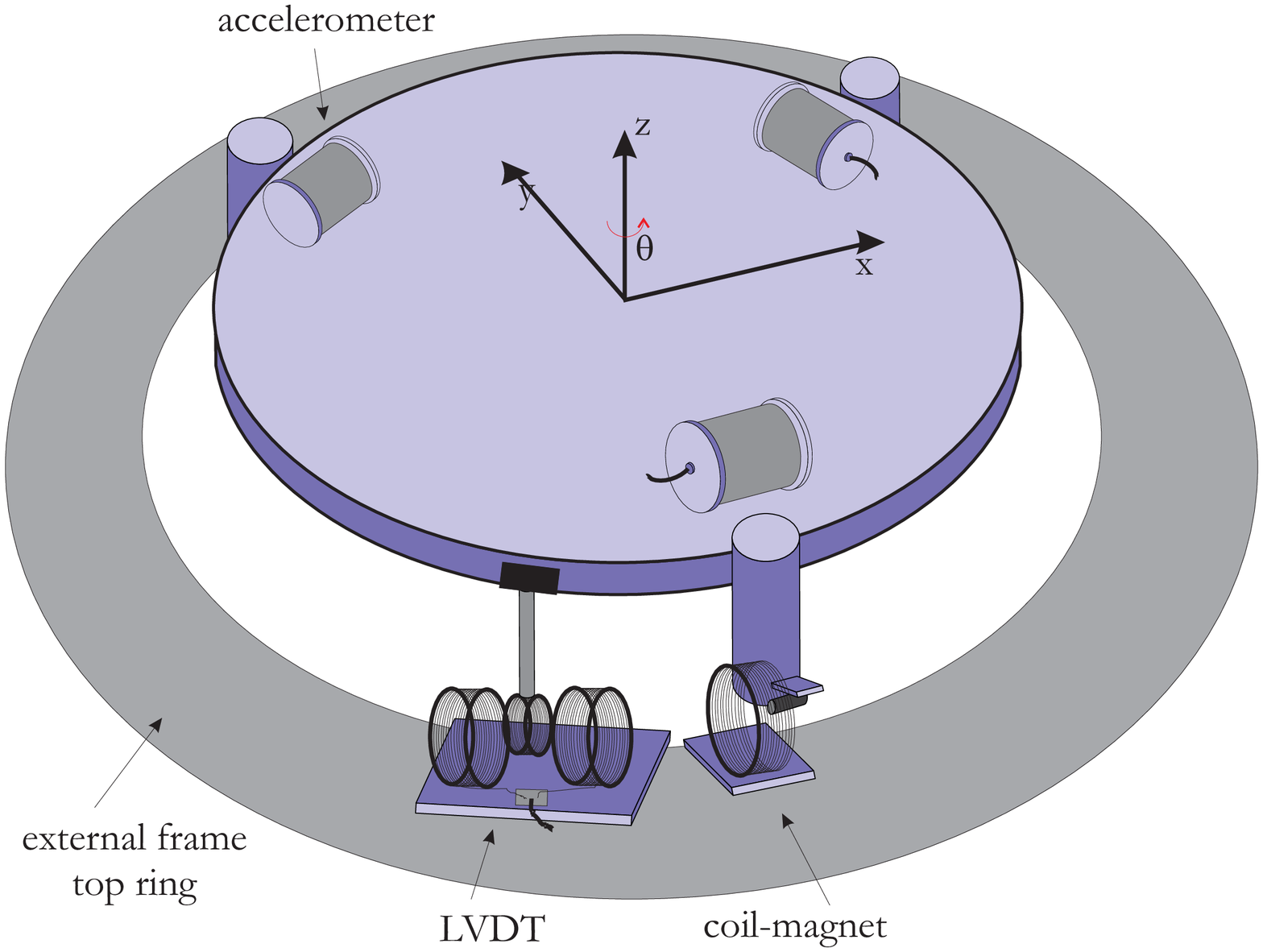}
\end{minipage}
\caption{\footnotesize LEFT: the superattenuator; RIGHT TOP:
logical scheme of the setup for the local active control; RIGHT
BOTTOM: simplified view of the IP top table, provided with the 3
accelerometers. One LVDT position sensors and one coil-magnet
actuator are also shown. } \label{setup} \enc\enf

\section{The approach to the problem of control}

The IP, where all the sensors are placed and on which the forces
are exerted, has 3 main resonant modes: two translation ($X,Y$)
and the rotation around the vertical axis ($\Theta$). Each sensor
is sensitive to all the modes and each actuator can excite all the
modes. In control theory language, such a system is defined MIMO
(multiple in-multiple out). Controlling a MIMO system can be very
difficult. Our approach has been different: the signals of the 3
sensors are digitally mixed (using proper transformation matrices)
to build up {\it virtual sensors}, sensitive to one mode only and
``blind" to the others. At the same time, we build up {\it virtual
actuators} able to excite each mode separately. The system is thus
uncoupled into 3 SISO (single in-single out) subsystems. In terms
of analytical mechanics this means the system is described in the
normal modes basis. The equations of motions take the form:
\beq \ddot{x}_k+\omega_k^2 x_k=q_k \;\;\;,\;\;\;k=1...3 \eeq
where $x_k$ is the $k-$th normal mode, $\omega_k/2\pi$ is the
corresponding resonant frequency and $q_k$ the generalized force
on that mode. Let ${\bf u}=(u_1,...,u_3)$ the vector made by the
outputs of the 3 sensors and let ${\bf x}=(x_1,...,x_3)$ the
vector made by the 3 virtual sensors. Analogously, let ${\bf v}$
the vector made by the 3 currents driving the actuators and ${\bf
q}$ the vector of the 3 generalized forces. The transformation
from the system of the physical sensors/actuators to that of the
virtual ones is operated by two matrices, such that:
\begin{eqnarray} {\bf v}&=&{\bf Dq} \label{vdq}\\ {\bf
x}&=&{\bf Su} \label{xsu} \end{eqnarray}
The {\it sensing} (${\bf S}$) and {\it driving} (${\bf D}$)
matrices are experimentally measured. The measurement procedure is
described in details in ref. 9. The result of the measurement is
determined by the mechanical characteristics of the system
(resonant frequencies, quality factors), the geometry of the
sensors and actuators and their calibration, but it is not needed
to know them in order to measure the matrices. The logic of
diagonalisation is explained by fig. \ref{decomp}.

After the diagonalisation the system can be considered as composed
of 3 {\it uncoupled} oscillators, and the control strategy for
each of them has to be defined independently (see fig. \ref{xth}).
\bef\bec \ingr[width=.6\textwidth]{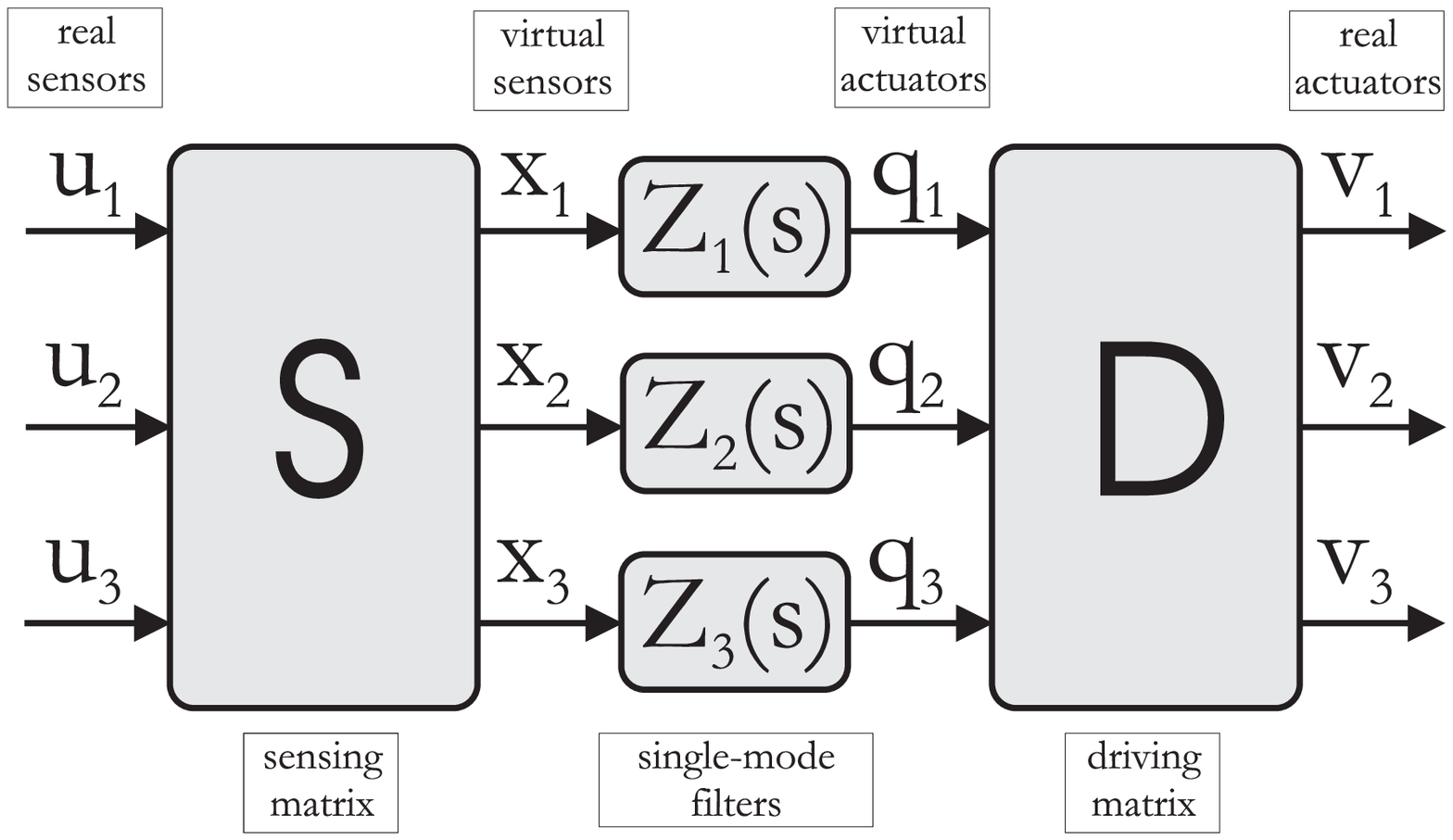}
\caption{\footnotesize The logic of the diagonalisation: the
output $u_i$ of the sensors are linearly combined by a matrix {\bf
S} in order to produce 3 {\it virtual} sensors outputs ($x_i$),
sensitive to pure modes. Three independent feedback filters
$Z_i(s)$ are designed for the pure modes and 3 generalized forces
$q_i$ are produced. The $q_i$ are turned into real currents
($v_i$) to feed the actuators via the matrix {\bf D}.}
\label{decomp} \enc\enf \bef\bec
\begin{minipage}{.47\textwidth}
\ingr[width=.95\textwidth]{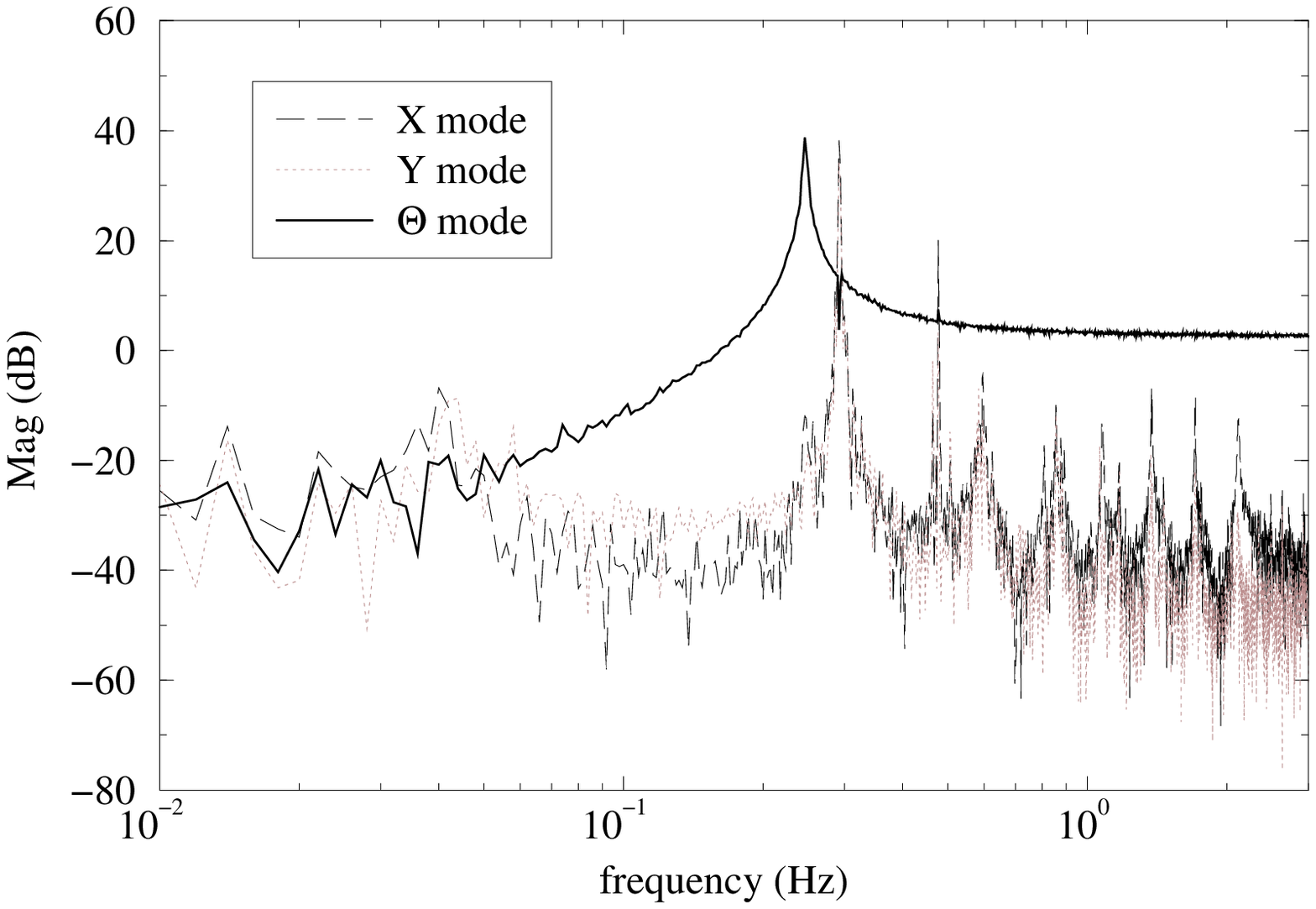}
\end{minipage}\hfill
\begin{minipage}{.47\textwidth}
    \includegraphics[width=\textwidth]{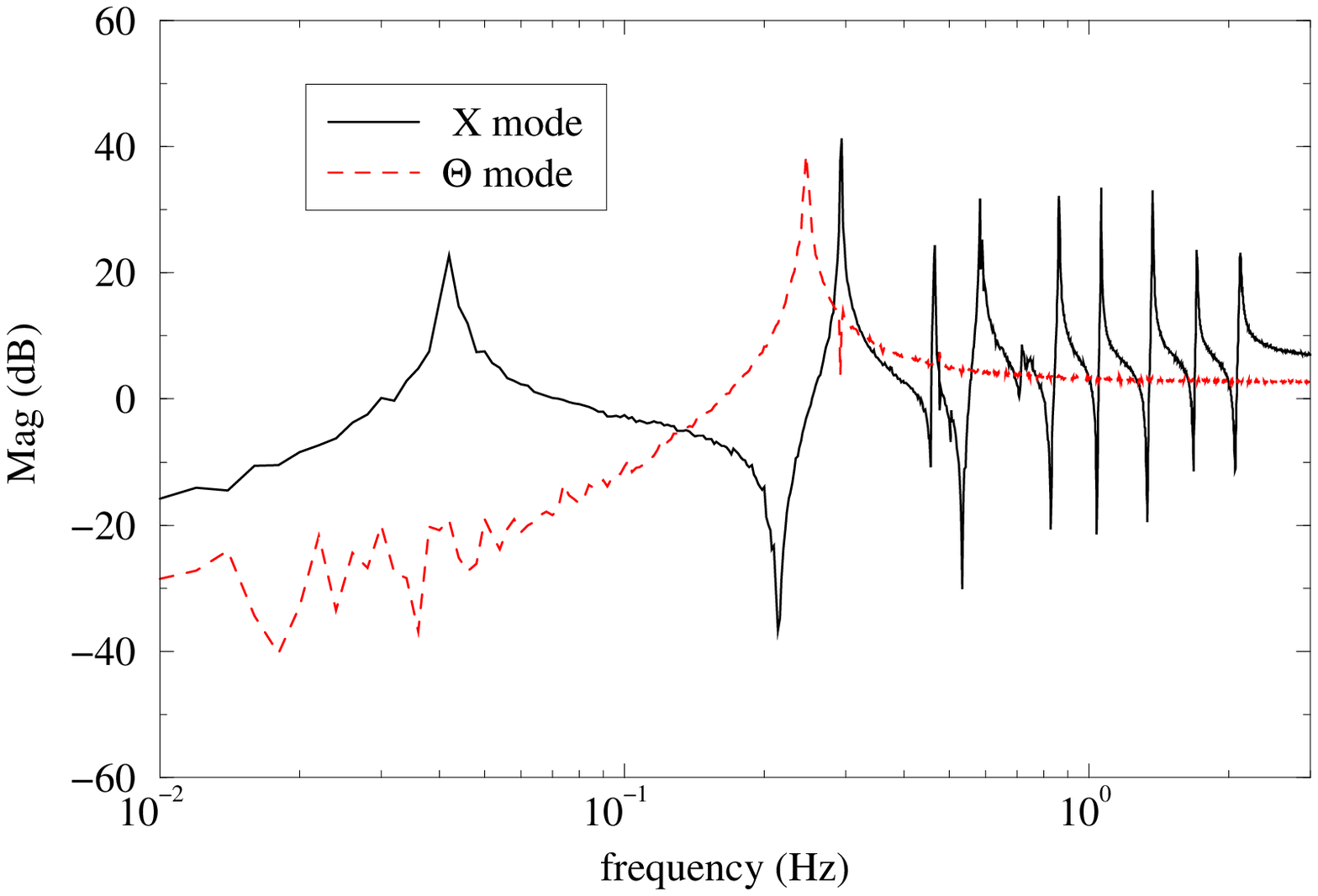}
\end{minipage} \caption{\footnotesize Effect of the digital diagonalisation.
LEFT: the output of the 3 virtual accelerometers when $\Theta$ is
excited. RIGHT: the output of the virtual accelerometers
    $X$ and $\Theta$ are compared. Different feedback strategies are
    required in the two cases, because $X$ senses all the translational
    modes of the SA chain. }\label{xth}
     \enc\enf

\section{Inertial damping: principle}

The control we describe here is called {\it inertial damping}
because it is performed by using (mostly) {\it inertial sensors}
(accelerometers). In the following , with the help of a simple
model, we explain why this is the best choice to achieve a high
performance damping.

Let us consider a simple pendulum of mass $m$ and length $l$. Let
$x$ be the abscissa of the suspended mass, $x_0$ that of the
suspension point. Let $F_{\rm fb}$ the external force on the
pendulum (i.e. the feedback force to control it). The equation of
motion is then:
\beq F_{\rm fb}=m\ddot{x}+\gamma\dot{x}+k(x-x_0) \label{eqschema}
\eeq
where $\gamma$ is the viscous dissipation factor and $k=mg/l$. The
control loop of such a system is sketched in fig. \ref{schema},
where $H(s)$ is the mechanical transfer function, $G(s)$ is the
compensator and {\it out} is the output of the sensor used. The
goal of the control is to damp the pendulum resonance. This can be
done easily with a {\it viscous} (theoretical) feedback force:
\beq F_{\rm fb}=-\gamma'\dot{x} \eeq
Our sensors do not measure $x$. Their output is:
\beq {\mathit out}=\left\{\begin{array}{ll} x-x_0 &\;\; \mbox{for
displacement sensors} \\ \ddot{x} &\;\; \mbox{for accelerometers}
\end{array} \right. \eeq
\bef\bec \ingr[width=.8\textwidth]{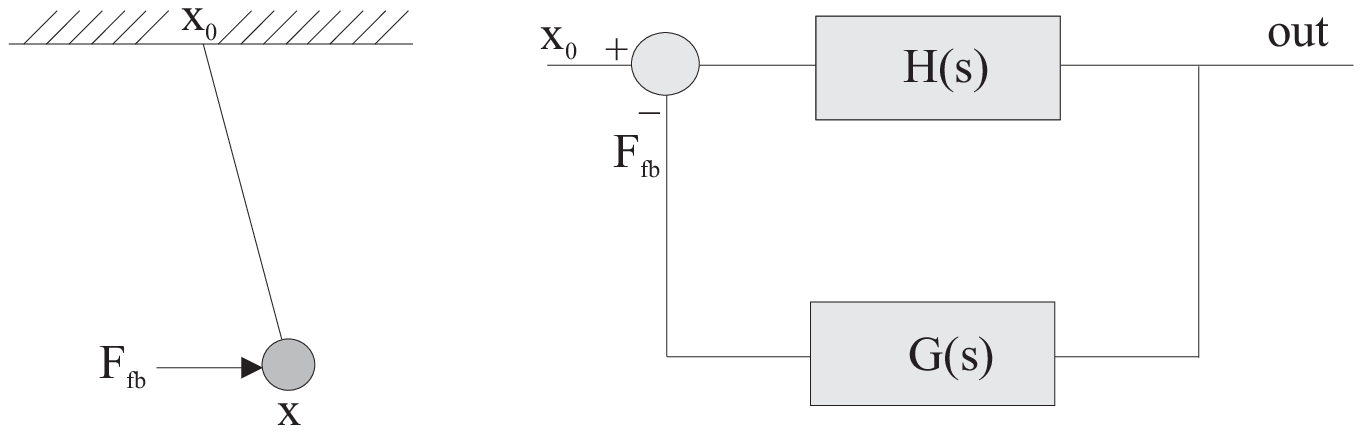} \caption{The control
scheme for a simple pendulum} \label{schema}\enc\enf
Therefore, the actual ``viscous" force that can be built if
position sensors are used has the form:
\beq F^p_{\rm fb}=-\gamma'\frac{\rm d}{{\rm d}t}(x-x_0) \eeq
It can be easily shown that with such a feedback force the closed
loop equation of motion (in Laplace space) reduces to:
\beq
x(s)=\frac{\omega_0^2+G_0s}{s^2+\omega_0^2+(\omega_0/Q+G_0)s}\cdot
x_0(s)  \label{pcltf} \eeq
where $G_0=\gamma'/m$ is a gain parameter\footnote{In a real
feedback system a frequency dependent gain function $G(s)$ rather
than a gain parameter has to be considered.} measuring the
intensity of the viscous feedback force, and $Q$ is the open loop
quality factor. When the loop is closed a damping of the resonance
is achieved:
\beq Q'\stackrel{G>>1}{\longrightarrow}\frac{\omega_0}{G} \eeq
Nevertheless, as the gain is increased, a larger amount of noise
is reinjected off-resonance. This is associated to the term
``$G_0s$" in the numerator of (\ref{pcltf}) and depends on the
fact that the sensor used to build up the feedback force measures
the position of the pendulum with respect to ground. Therefore, an
infinitely efficient feedback would ``freeze" the pendulum to
ground (which is seismic noisy), reducing its motion at the
resonance, with the drawback of bypassing its attenuation
properties above resonance.

The situation is fairly different when an inertial sensor is used.
In this case the viscous feedback force is obtained by integrating
the accelerometer output, and the output does not depend on $x_0$:
\beq F^a_{\rm fb}=-\gamma'\int \ddot{x}{\rm d}t \eeq
The closed loop equation of motion is then:
\beq x(s)=\frac{\omega_0^2}{s^2+\omega_0^2+(\omega_0/Q+G_0)s}\cdot
x_0(s)  \label{acltf} \eeq
A damping of the resonance is obtained (exactly as in the previous
case) but without reinjection of off-resonance noise. In fig.
\ref{apcomp} a simulation of the closed loop transfer function
$x(s)/x_0(s)$ is shown in the two cases.
\bef\bec
\begin{minipage}{.47\textwidth}
\ingr[width=\textwidth]{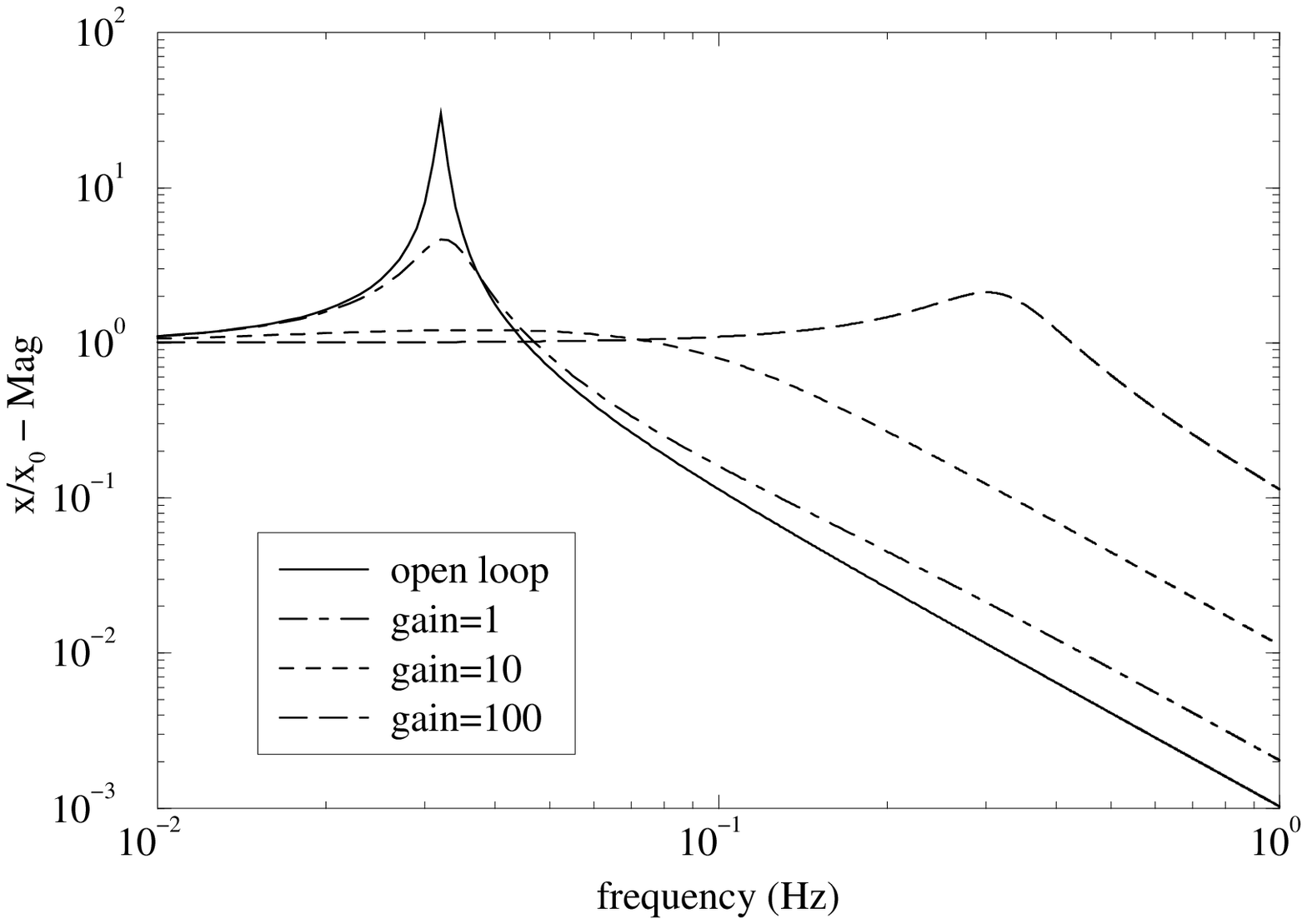}
\end{minipage}\hfill
\begin{minipage}{.47\textwidth}
\ingr[width=\textwidth]{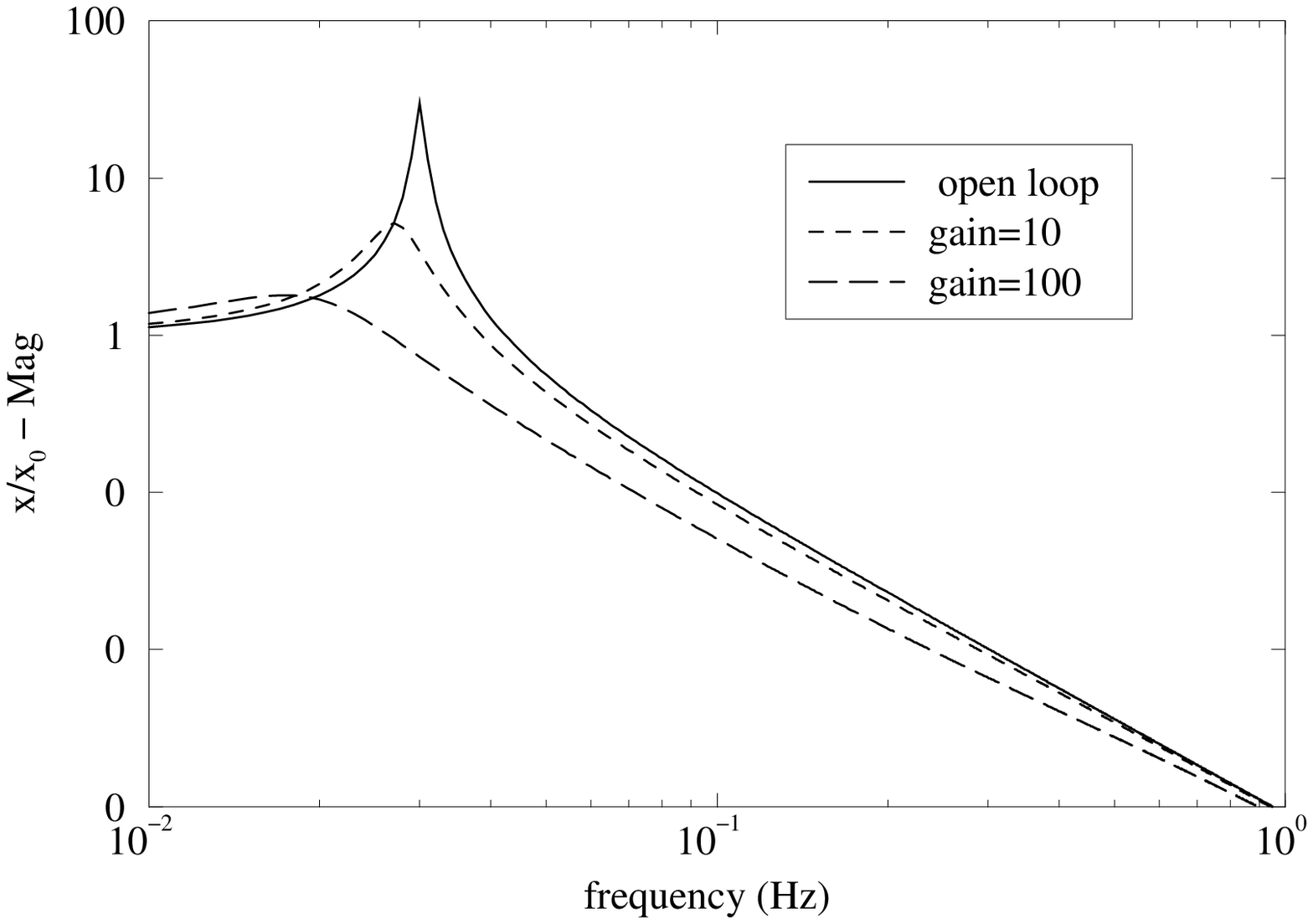}
\end{minipage}
\caption{Damping of a simple pendulum: the closed loop transfer
function $x(s)/x_0(s)$ (magnitude) when a position sensor is used
(LEFT) and when an accelerometer is used (RIGHT).} \label{apcomp}
\enc\enf
Up to now we have considered a simple viscous damping. It is
possible to increase the bandwidth of the control if the feedback
force contains a term proportional to $x$ (the double integral of
the accelerometer signal). The result obtained in this case is
shown in fig. \ref{xdamp}.
 \bef\bec
\begin{minipage}{.47\textwidth}
 \includegraphics[width=\textwidth]{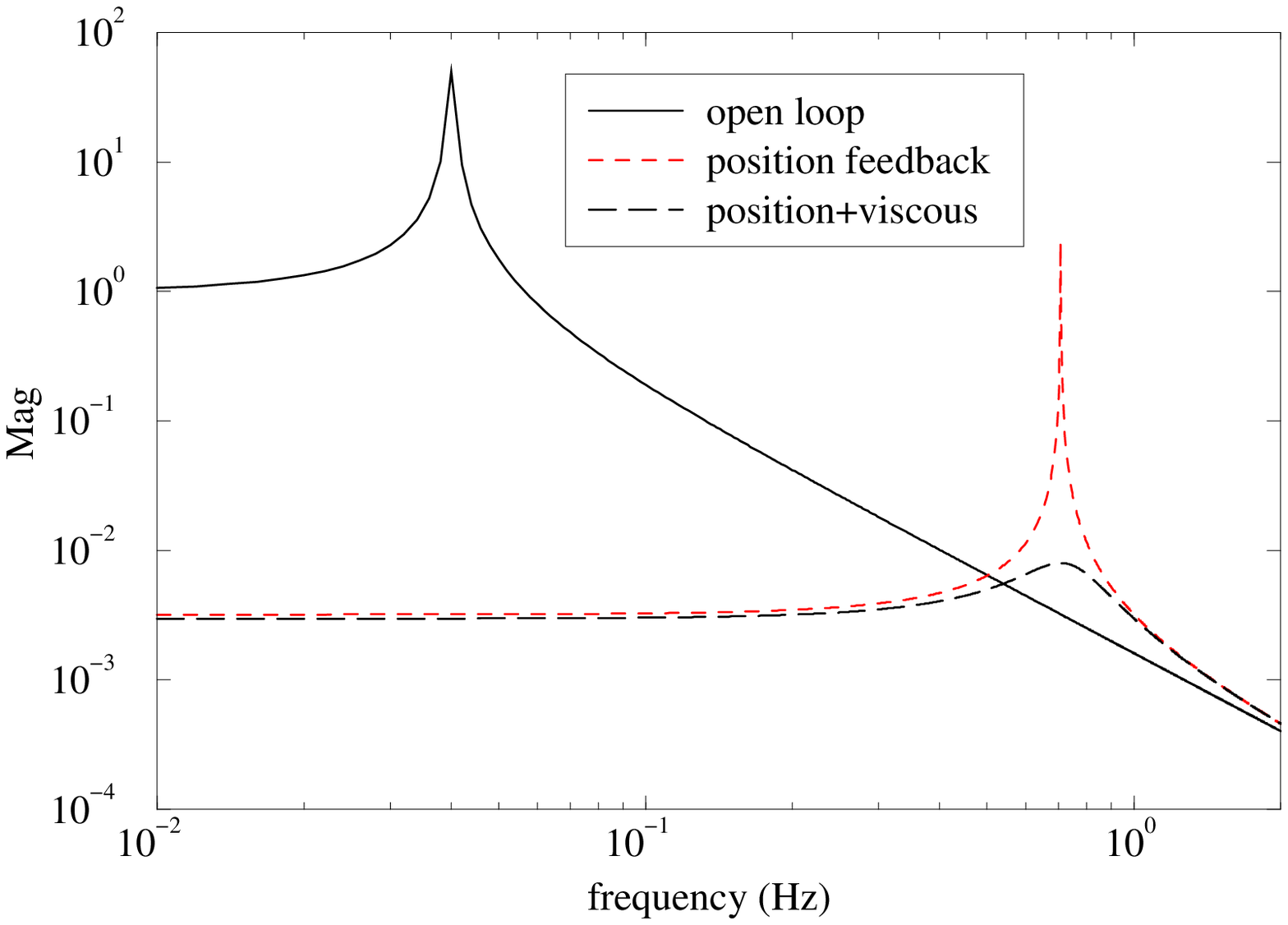}
\caption{\footnotesize Inertial damping of a simple pendulum when
a position feedback is implemented. } \label{xdamp}\end{minipage}
\hfill
\begin{minipage}{.47\textwidth}
 \includegraphics[width=\textwidth]{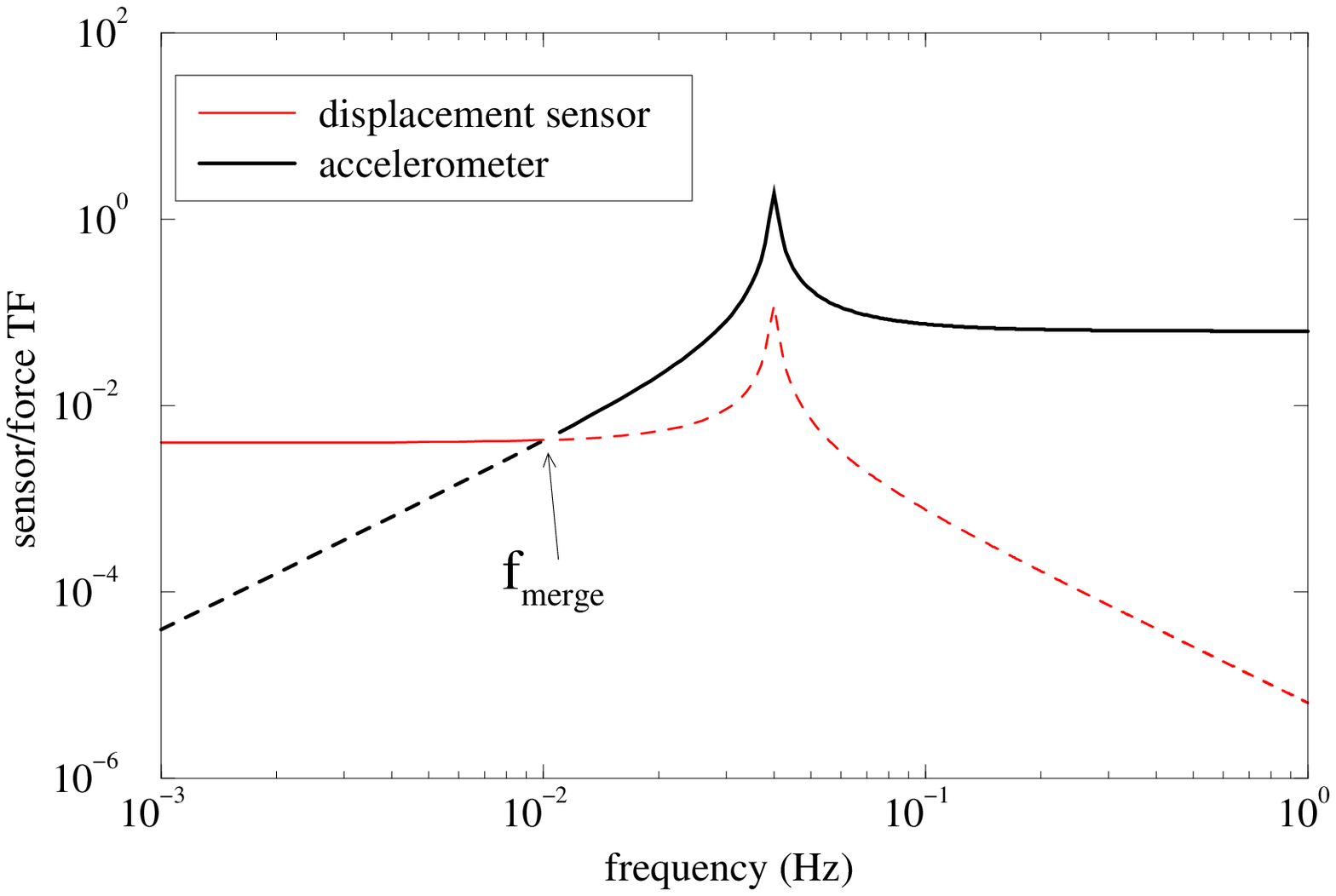}
\caption{\footnotesize {\it Merging} of displacement and
acceleration sensors (simulation for a simple pendulum).}
\label{merge}
\end{minipage}
\enc\enf

\section{Control strategy}

In this section we extend the principles of the previous section
and describe the strategy to control the SA.

The basic idea of inertial damping is to use the accelerometer
signal to build up the feedback force. As a matter of fact, an
infinitely efficient feedback using only the inertial sensor
information, would null the acceleration of the pendulum, but it
would not do anything if the pendulum moves at constant velocity.
Such a control would be unstable with respect to {\it drifts}.
Therefore, if the control band is to be extended down to DC, a
position signal is necessary. Our solution was a {\em merging} of
the two sensors: the virtual LVDT (position) and accelerometer
signals are combined in such a way that the LVDT signal ($l(s)$)
dominates below a chosen cross frequency $f_{\rm merge}$ while the
accelerometer signal ($a(s)$) dominates above it (see fig.
\ref{merge} and ref. 7). The feedback force has the
form\footnote{Actually, the LVDT signal $l(s)$ is properly
filtered in order to preserve feedback stability at the crossover
frequency and in order to reduce the amount of reinjected noise at
$f>f_{\rm merge}$.}:
\beq F_{\rm fb}=G(s)\left[a(s)+\epsilon l(s)\right] \eeq
where $G(s)$ is the digital filter transfer function (see fig.
\ref{filter}) and $\epsilon$ is the parameter whose value
determines $f_{\rm merge}$. We have chosen $f_{\rm merge}\sim 10$
mHz (corresponding to $\epsilon\sim 5\cdot 10^{-3}$). This
approach stabilizes the system with respect to low frequency
drifts at the cost of reinjecting a fraction $\epsilon$ of the
seismic noise via the feedback.

We describe in the following the feedback design for the 3 d.o.f.,
starting from the the translational ones. The virtual $X$ and $Y$
sensors show many resonant peaks (the modes of a chain of
pendulums) and this requires a more sophisticated feedback
strategy. The digital filter used to control the translation modes
($G(s)$) is shown in fig. \ref{filter} (LEFT). It shows three main
features:
\begin{itemize}
\item for $0.01<f<2$ Hz the gain is proportional to $f^{-2}$. This
corresponds to the case of fig. \ref{xdamp}: the accelerometer
signal is integrated twice and the feedback force is proportional
to $x$;
\item for $f>2$ Hz the gain is proportional to $f^{-1}$. The
accelerometer signal is integrated once: the feedback force is
proportional to the velocity and a viscous damping is achieved;
\item the peaks visible in the filter are necessary to compensate
the corresponding dips in the mechanical transfer function
($H(s)$) of fig. \ref{xth}, in order to make the feedback stable.
\end{itemize}
Fig. \ref{filter} (RIGHT) shows the open loop gain transfer
function $G(s)H(s)$.

The damping strategy for the $\Theta$ mode is simpler: the
$\Theta$ virtual sensor (fig. \ref{xth}, RIGHT) shows one
resonance peak only and no dips: no compensation is necessary.
Apart from this, the feedback strategy is similar to the ones used
for the translational modes.
\bef\bec
\begin{minipage}{.47\textwidth}
\ingr[bb=130 60 693 473,clip=true,width=\textwidth]{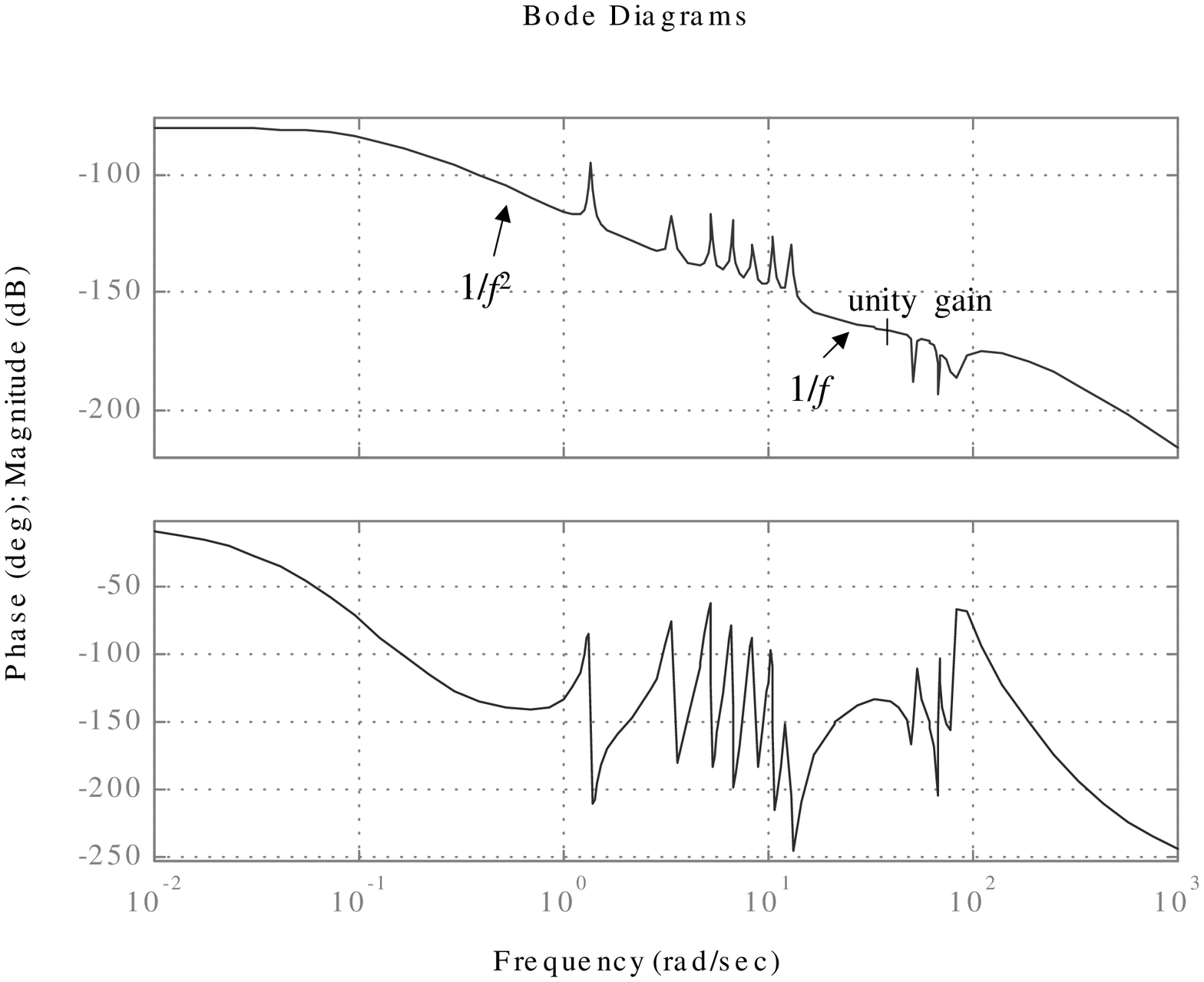}
\end{minipage}\hfill
\begin{minipage}{.47\textwidth}
\ingr[width=\textwidth]{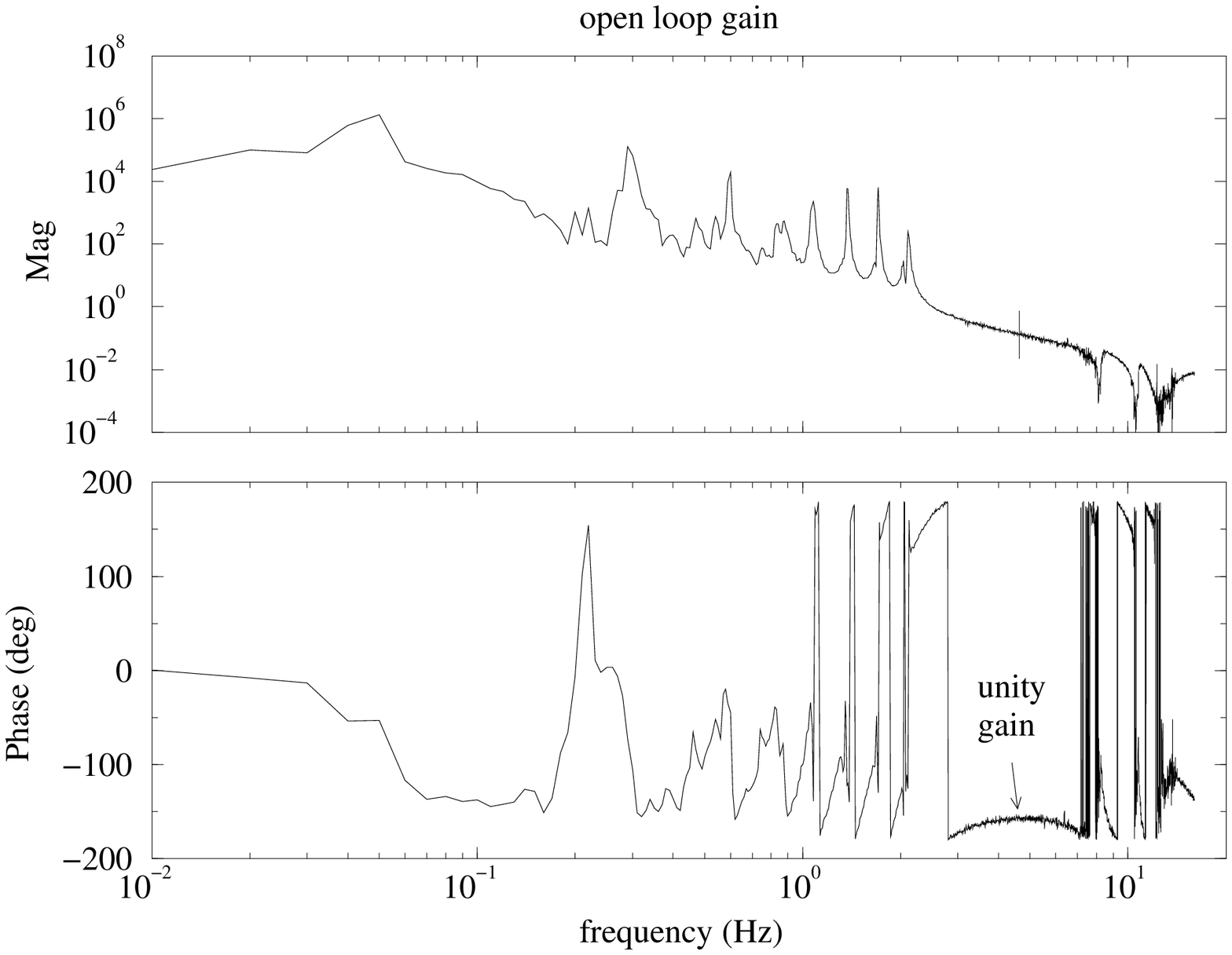}
\end{minipage}
\caption{\footnotesize LEFT: Digital filter used for the inertial
damping of a translation mode ($X$). The filter slope is $f^{-2}$
in the range 10 mHz$<f<$3 Hz, $f^{-1}$ for $f>$3 Hz. The unity
gain is at 4 Hz. The peaks in the digital filter are necessary to
compensate the dips in the mechanical transfer function (see the
transfer function of the $X$ mode in fig. \ref{xth}). RIGHT: open
loop gain function (measured). The phase margin at the unity gain
frequency is about $25^\circ$.} \label{filter} \enc\enf
\section{Inertial damping: experimental results}

The result of the inertial control (on 3 d.o.f.) is shown in
figure \ref{results}. The measurement has been performed in air.
The noise on the top of the IP is reduced over a wide band (10 mHz
- 4 Hz). A gain $>1000$ is obtained at the main SA resonance (0.3
Hz). The RMS translational motion of the IP (calculated as $x_{\rm
RMS}(f)=\sqrt{\int_f^\infty \tilde{x}^2(\nu){\rm d}\nu}$) in 10
sec. is reduced from more than 30 to 0.3 $\mu$m. The closed loop
floor noise corresponds to the fraction of seismic noise
reinjected by using the position sensors for the DC control and
can, in principle, be reduced by a steeper low pass filtering of
the LVDT signal at $f>f_{\rm merge}$ and by lowering $f_{\rm
merge}$: both this solution have drawbacks and need a careful
implementation.

Preliminary measurements of the displacement of the mirror with
respect to ground have been performed in air, using an LVDT
position sensor. The residual RMS mirror motion in 10 sec.
is\footnote{This number has been obtained with a simpler feedback
design, less {\it aggressive} than the one of fig. \ref{filter}:
the gain raised as $1/f$ (pure viscous damping force), the
crossover frequency was 30 mHz and no compensation of the dips was
needed.}: \beq x_{\rm RMS}(0.1\,{\rm Hz})\leq 3\;\mu{\rm m}. \eeq
When the damping is on such a measurement can provide only an
upper bound because the LVDT output is dominated by the seismic
motion of the ground.

\bef\bec
\begin{minipage}{.47\textwidth}
    \includegraphics[width=\textwidth]{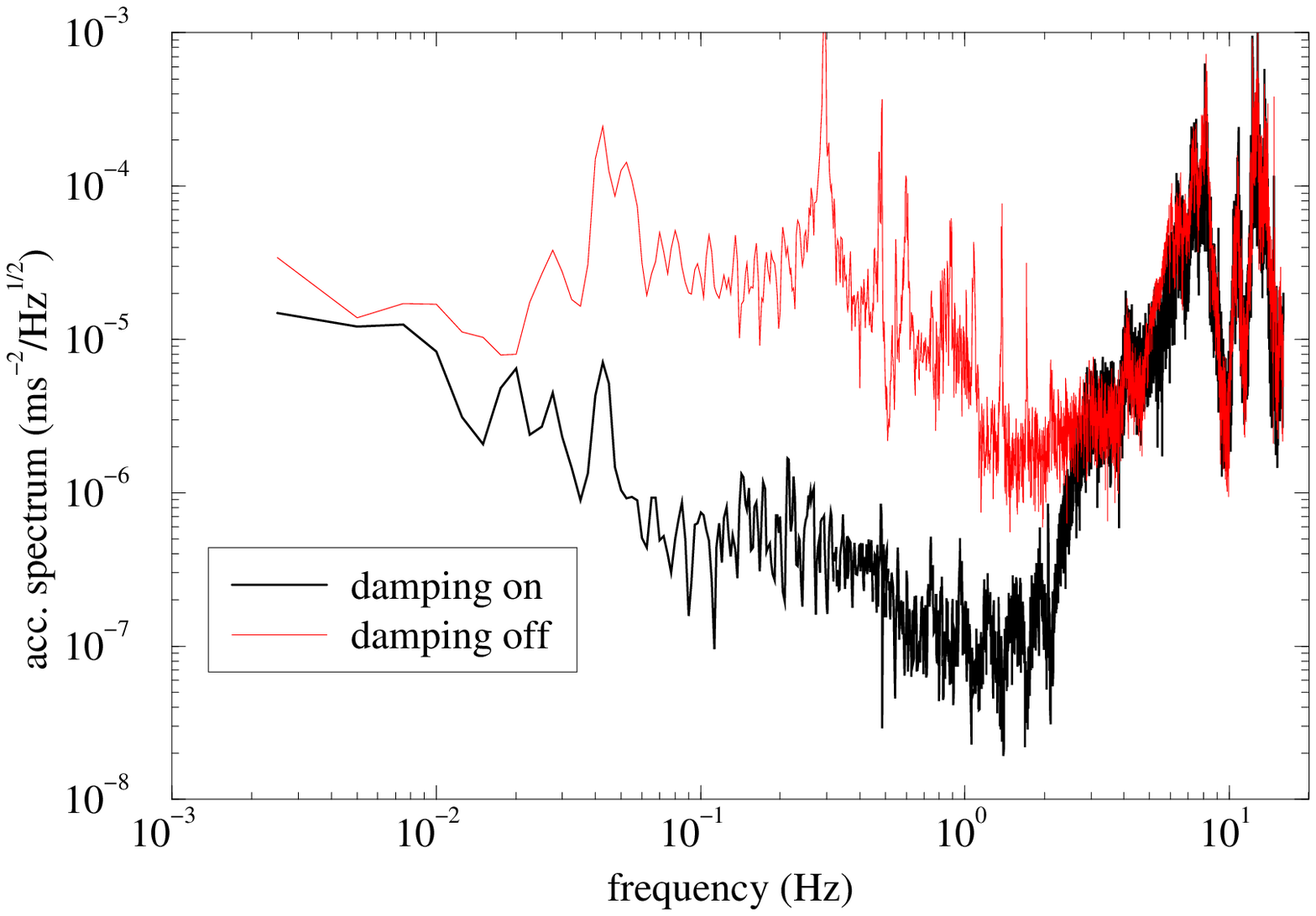}
\end{minipage}\hfill
\begin{minipage}{.47\textwidth}
 \includegraphics[width=\textwidth]{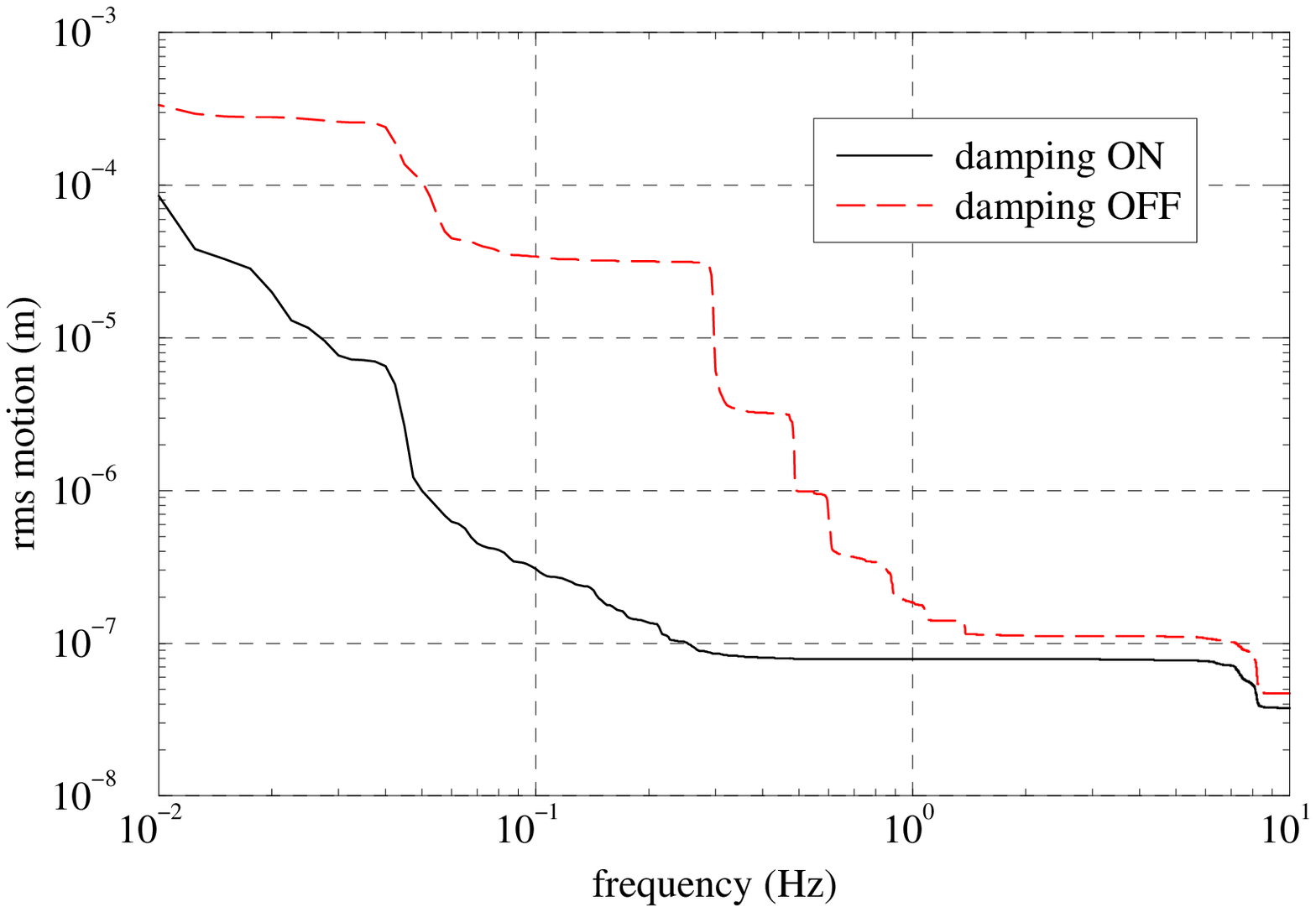}
\end{minipage}
\caption{\footnotesize Performance of the inertial control
($X,Y,\Theta$ loops closed) of the superattuenuator, measured on
the top of the IP: the left plot shows the acceleration spectral
density as measured by the {\it virtual} accelerometer $X$
(translation). The right plot shows the effect of the feedback on
the RMS residual motion of the IP as a function of the frequency.
} \label{results} \enc\enf

\section{Conclusions}

In this lecture we have tried to outline how to face the problem
of reducing the free motion of the suspended optical components of
the VIRGO interferometer, associated to the resonances of the
suspension. This is only the beginning: once the motion of the
mirrors is reduced to a few microns, the lower control stages can
operate to {\it lock} the interferometer in the correct operation
state.

\section*{Acknowledgments}

The author wishes to thank all the people of the Pisa and Florence
VIRGO Groups. Among them, special thanks to Diego Passuello,
Alberto Gennai and Andrea Marin.

\end{document}